\documentclass[aps, preprint, prc, amsmath, amssymb, nofootinbib, superscriptaddress, showkeys]{revtex4-1}
\usepackage{subcaption}
\usepackage{booktabs}
\usepackage{braket}
\usepackage{enumitem}
\usepackage{slashed}
\usepackage{mathrsfs}
\usepackage{multirow}
\usepackage{graphicx}
\usepackage{dcolumn}
\usepackage{bm}
\usepackage{comment}
\usepackage{tikz}
\usepackage{ulem}
\usepackage{CJK}
\usepackage[absolute,overlay]{textpos}
\usetikzlibrary{patterns}
\usetikzlibrary{decorations.markings}
\usetikzlibrary{decorations.pathmorphing, decorations.markings, arrows}
\tikzset{
  arrow at/.style={
    decoration={
      markings,
      mark=at position #1 with {\arrow{latex}}
    },
    postaction={decorate}
  }
}
\usepackage{pgfplots}
\pgfplotsset{compat=1.18}

\usepackage{hyperref}
\usepackage{comment}
\newcommand{\onbb}{$0\nu\beta\beta$}

\allowdisplaybreaks

\begin{document}

\begin{CJK*}{UTF8}{gbsn}

\title{The lepton-number-violating pion decay and the type-\uppercase\expandafter{\romannumeral 1} seesaw mechanism in chiral perturbation theory}

\author{You-Cai Chen (陈友才)}
\affiliation{Jilin University, Changchun 130012, China}

\author{Tai-Xing Liu (刘太兴)}
\email{txliu@impcas.ac.cn}
\affiliation{Institute of Modern Physics, Chinese Academy of Sciences, Lanzhou, 730000, China}
 
\author{Dong-Liang Fang (房栋梁)}
\affiliation{Institute of Modern Physics, Chinese Academy of Sciences, Lanzhou, 730000, China}
\affiliation{School of Nuclear Science and Technology, University of Chinese Academy of Sciences, Beijing 100049, China}

\begin{abstract}

We investigate the process of lepton-number-violating pion decay, which dominates the nuclear neutrinoless double beta decay induced by the short-range operator, within the type-\uppercase\expandafter{\romannumeral1} seesaw mechanism. The type-\uppercase\expandafter{\romannumeral1} seesaw mechanism gives rise to the Dirac and Majorana mass terms of neutrinos by introducing the gauge-singlet right-handed neutrinos, which are usually called sterile neutrinos. Using chiral perturbation theory, the transition amplitudes in the case of the light and heavy sterile neutrinos are calculated up to $\mathcal{O}(Q^2/\Lambda^2_\chi)$ respectively, where $Q$ is the typical low-energy scale in this process and $\Lambda_\chi$ the chiral symmetry breaking scale. We then adopt a naive interpolation formula of mass dependence to obtain the amplitude in the full mass range and briefly discuss its validity.

\end{abstract}


\begin{textblock}{12}(1.55, 15)
\textbf{\footnotesize Submitted to Chinese Physics C}
\end{textblock}

\maketitle

\end{CJK*}

\section{Introduction}
The charged fermions in the Standard Model of particle physics (SM) acquire mass via its Yukawa coupling with the Higgs field after the breaking of electroweak symmetry. The SM predicts massless neutrinos due to the absence of right-handed neutrinos. However, the observed neutrino oscillation for two decades~\cite{Super-Kamiokande:1998kpq, SNO:2001kpb, K2K:2002icj} convincingly confirms that the masses of neutrinos are non-zero. The simplest way to accommodate the neutrinos with nonvanishing mass is the well-known type-\uppercase\expandafter{\romannumeral1} seesaw mechanism~\cite{Minkowski:1977sc, Yanagida:1979as, Gell-Mann:1979vob, Glashow:1979nm, Mohapatra:1979ia}, or so-called neutrino-extended SM ($\nu$SM),  which introduces two or more right-handed neutrino fields $\nu_R$ manifested as gauge singlet under the SM gauge transformation such that the SM gauge symmetry is preserved. Besides explaining the origin of the tiny masses of active neutrinos, the sterile neutrinos can also be utilized to explain the baryon asymmetry of the universe and dark matter~\cite{Asaka:2005pn, Shaposhnikov:2008pf}. Another attractive feature is that the type-\uppercase\expandafter{\romannumeral1} seesaw mechanism can naturally capture the Majorana neutrinos that induce the neutrinoless double beta ({\onbb}) decay~\cite{Schechter:1981bd}.

As a promising probe of the lepton number violation (LNV), the hadronic {\onbb} decay plays an important role in particle and nuclear physics. For this process, both the GERDA~\cite{GERDA:2020xhi} and KamLAND-Zen~\cite{KamLAND-Zen:2022tow} projects, which use $^{76}\text{Ge}$ and $^{136}\text{Xe}$ isotopes respectively, provide the current best limits on the half-lives up to $10^{26}$ years. In the future, ton-scale facilities are expected to improve this sensitivity by one or two orders of magnitude~\cite{LEGEND:2017cdu, nEXO:2017nam}. On the theoretical side, an effective field theory (EFT) roadmap for matching various LNV models at high energy to the nuclear operators at low energy has been accomplished in Refs.~\cite{Cirigliano:2017djv, Cirigliano:2018yza}. Starting with the chiral EFT framework, the {\onbb} nuclear matrix elements (NMEs) of the nuclei relevant for experiments are available with \textit{ab initio} nuclear many-body methods~\cite{Yao:2019rck, Belley:2020ejd, Novario:2020dmr, Belley:2023lec}. Despite significant progress, accurately predicting NMEs remains a substantial challenge because of the nonperturbative effects in nuclear systems (for a comprehensive review see Ref.~\cite{Agostini:2022zub}).

Compared with the nuclear {\onbb} decay, the process involving only mesons can be treated in a perturbative manner, providing a simpler way to probe the LNV mechanism (For an example see Refs.~\cite{Liao:2019gex, Liao:2020roy, Zhou:2021lnl}). Given that the transition $\pi^-\pi^-\to e^-e^-$ (or unphysical transition $\pi^-\to\pi^+e^-e^-$) dominates the nuclear {\onbb} decay induced by the short-range LNV mechanisms, such as R-parity violation supersymmetry mechanism~\cite{Faessler:1996ph}, we explore this transition within the chiral perturbative theory (ChPT). Moreover, Refs.~\cite{Feng:2018pdq, Tuo:2019bue} (Ref.~\cite{Nicholson:2018mwc}) calculated the long-range (short-range) contribution to the low-energy transition with the lattice QCD method. Such results can be used to determine the low-energy constants (LECs) that appear in the amplitude for the long- and short-distance mechanisms, respectively.

The intriguing features of the type-\uppercase\expandafter{\romannumeral1} seesaw mechanism prompt many investigations into the impact of sterile neutrinos on the {\onbb} decay rate~\cite{Liao:2019gex, Liao:2020roy, Zhou:2021lnl, Faessler:2014kka, Dekens:2020ttz, Fang:2021jfv, Dekens:2023iyc, Fang:2024hzy, Dekens:2024hlz}. Most of the works concentrate on the tree-level effects of sterile neutrinos on the hadronic decay rate. In this work, we move the calculation of the process $\pi^-\pi^-\to e^-e^-$ in the context of the type-\uppercase\expandafter{\romannumeral1} seesaw mechanism to one-loop level, corresponding to $\mathcal{O}(Q^2/\Lambda_\chi^2)$ in ChPT. Introducing sterile neutrinos leads to a mass-dependent description and consequently imposes more constraints on {\onbb} decay. In light of the competition between the sterile neutrino mass with the breakdown scale $\Lambda_\chi$, as is commonly done, we integrate the neutrinos with mass above $\Lambda_\chi$ out at the quark level, whereas the neutrinos with mass below $\Lambda_\chi$ are kept as explicit degrees of freedom in ChPT. We note that a description of this process within the standard mechanism of SM neutrino of {\onbb} decay is presented up to the one-loop level by Ref.~\cite{Cirigliano:2017tvr}.

This paper is organized as follows. In Sec.\ref{sec:theory}, we present the theoretical formalism of the {\onbb} amplitude and the interpolation formula. This is followed by results and discussions in Sec.\ref{sec:resluts}. Finally, we summarize in Sec.\ref{sec:summary}.

\section{theory framwork}\label{sec:theory}
\subsection{Sterile Neutrinos in the $\nu$SM}

The central idea of the type-\uppercase\expandafter{\romannumeral 1} seesaw mechanism to address the shortcomings of SM is to introduce sterile neutrinos. In this model, the effective Lagrangian consists of the SM Lagrangian and its extension based on the $k$ right-handed neutrino fields $N_{Ri}$ :
\begin{equation}
    \mathcal{L}=\mathcal{L}_\text{SM}-\left[\overline{\psi_L}\widetilde{H}Y_DN_R+\frac{1}{2}\overline{N}_R^cM_RN_R+h.c.\right],
\end{equation}
where $\psi_L=(\nu_{L},e_L)^T$ denotes the left-handed lepton doublet, $\widetilde{H}=i\tau_2H^*$ the conjugate state of the Higgs doublet $H$. $Y_D$ is a $3\times k$ Yukawa coupling matrix and $M_R$ a symmetric $k\times k$ matrix. The charge conjugation of the fermion field is defined as $\psi^c=C\bar{\psi}^T$ in terms of the charge conjugation operator $C=i\gamma_2\gamma_0$. After the electroweak symmetry breaking, the neutrinos acquire masses
\begin{equation}
    \mathcal{L}_\text{m}=-\frac{1}{2}(\overline{\nu}_{L},\overline{N}^c_R)\left(\begin{matrix}
        0&&M_D\\
        M_D^T&&M_R
    \end{matrix}\right)\left(\begin{matrix}
        \nu_L\\
        N^c_R
    \end{matrix}\right)+h.c.,
\end{equation}
where $M_D=\frac{v}{\sqrt{2}}Y_D$ denotes the Dirac mass, in which $v=246$ GeV is the Higgs vacuum expectation value. Using the unitary mixing matrix $\mathcal{U}$, one can diagonalize the $(3+k)\times(3+k)$ mass matrix:
\begin{align}
    \mathcal{U}^T\left(\begin{matrix}
        0&&M_D\\
        M_D^T&&M_R
    \end{matrix}\right)\mathcal{U}=\left(\begin{matrix}
        \hat{M}_\nu&&0\\
        0&&\hat{M}_N
    \end{matrix}\right),\,\,\,\,\mathcal{U}=\left(\begin{matrix}
        U&&R\\
        S&&V
    \end{matrix}\right),
    \label{diageq}
\end{align}
where $\hat{M}_\nu=\text{diag}(m_1, m_2, m_3)$ is the mass matrix of the three generations of the active neutrino, $\hat{M}_N=\text{diag}(M_1, M_2, ..., M_k)$ the mass matrix of the $k$ flavors of sterile neutrinos. $U$ is the so-called light neutrino mixing matrix, i.e. the PMNS matrix~\cite{Pontecorvo:1957qd, Maki:1962mu}. One should note that the $U$ matrix is no longer unitary due to the existence of sterile neutrinos. $R$, a $3\times k$ matrix, represents mixing the sterile neutrinos in the charged current (CC). To be specific, the neutrino flavor eigenstate $\nu_\alpha$ $(\alpha= e, \mu, \tau)$ is related to the neutrino mass eigenstate via
\begin{equation}
    \nu_\alpha=U_{\alpha i}\nu_{i}+R_{\alpha j}N_j.
    \label{neutmix}
\end{equation}
In the context of the Majorana neutrinos, the propagators can be expressed as~\cite{Doi:Masaru}
\begin{equation}
\begin{split}
    \bra{0} T\nu_i(x)\nu^T_j(y)\ket{0}&=iS_F(x-y)C^T\delta_{ij};\\
   \bra{0} TN_i(x)N^T_j(y)\ket{0}&=iS_F(x-y)C^T\delta_{ij};\\
   \bra{0} TN_i(x)\nu^T_j(y)\ket{0}&=0;
\end{split}
\end{equation}
where $S_F(x-y)$ is the standard fermion propagator.

Additionally, according to Eq.~\eqref{diageq}, such model requires the neutrino masses and mixing matrix elements to satisfy the relation:
\begin{equation}
\sum_iU^2_{ei}m_i+\sum_jR^2_{ej}M_j=0.
\label{sesarel}
\end{equation}
This seesaw relation implies that at least two sterile neutrinos must be in the mechanism. For the scenario with multiple sterile neutrinos, the amplitude results from the sum of the contributions of individual neutrinos, e.g. see Refs.~\cite{Fang:2021jfv, Fang:2024hzy}. Moreover, although Eq.\eqref{sesarel} imposes a constraint on the masses of sterile neutrinos, their specific values are model-dependent due to the existence of undetermined parameters. For example, $M_i\sim\mathcal{O}(10^{15})$ GeV arises naturally under the assumption $Y_D\sim\mathcal{O}(1)$, whereas keV-scale masses become viable through the soft $L_e-L_\mu-L_\tau$ flavor symmetry breaking~\cite{Lindner:2010wr}. Therefore, we focus on the mass dependence of the amplitude in this work to provide essential inputs adaptable to various models.

\subsection{Chiral Perturbation Theory for LNV pion decay}

To describe the effects of physics beyond SM much larger than the electroweak scale at a low-energy scale, a systematic EFT framework has been made in Refs.~\cite{Cirigliano:2017djv, Cirigliano:2018yza}. With this approach, the beyond-SM operators are first matched to a low-energy EFT (LEFT) with $SU(3)_c\times U(1)_\text{em}$ gauge symmetry via integrating out heavy SM particles whose masses are comparable to the electroweak scale such as Higgs particles. Further, the LEFT operators are evolved to the QCD scale using the renormalization group equations and then rewritten as chiral effective operators based on chiral symmetry. It is worth emphasizing that the degrees of freedom have been changed from quarks to hadrons (pion, nucleon, etc.) after matching LEFT onto ChPT due to the nonperturbative effects of QCD.

In the seesaw model, the ChPT description of a sterile neutrino depends on its mass $M_i$. For $M_i<\Lambda_\chi$, the (light) sterile neutrino as a degree of freedom in ChPT mediates the {\onbb} decay of pion. The relevant Lagrangian spelled by the pion fields $u=\text{exp}(i\pmb{\tau}\cdot\pmb{\pi}/(2F_0))$ and the leptonic charged current  $l^\mu=-2\sqrt{2}G_FV_{ud}\tau^+\,\overline{e_L}\,\gamma_\mu\nu_{eL}$ can be written as
\begin{equation}
    \mathcal{L}=\frac{F_0^2}{4}\text{Tr}\left[u^\mu u_\mu+u^\dagger\chi u^\dagger+u\chi^\dagger u\right],
\end{equation}
where $F_0$ is the pion decay constant in the chiral limit, the Fermi constant $G_F=1.166\times10^{-5}$ $\text{GeV}^{-2}$, the quark mixing matrix element $V_{ud}=0.97$, and
\begin{equation}
    \begin{split}
        &u_\mu=-i\left[u^\dagger(\partial_\mu-il_\mu)u-u\partial_\mu u^\dagger\right]\\
        &\chi=2B\,\text{diag}(m_u,m_d).
    \end{split}
\end{equation}
The constant $B$ ($\simeq$ 2.8 GeV) connects the leading order pion mass $M$ to the light quark masses via $M^2=B(m_u+m_d)$, called the quark condensate. For the sake of clarity, we expand the Lagrangian in the pion field and split it into two parts: the pure strong interaction
\begin{equation}
\begin{split}
\mathcal{L}_{\pi\pi}&=\frac{1}{6F^2_0}\left(2\pi_0\partial_\mu\pi_0+\pi^+\partial_\mu\pi^-+\pi^-\partial_\mu\pi^+\right)^2+\frac{M^2}{24F_0^2}\left(\pi^2_0+2\pi^+\pi^-\right)^2\\
&\ \ \ -\frac{1}{3F_0^2}(\pi_0^2+2\pi^+\pi^-)\left(\partial_\mu\pi_0\partial^\mu\pi_0+\partial_\mu\pi^+\partial^\mu\pi^-\right),
\end{split}
\end{equation}
describing the coupling between pion fields and the CC operators
\begin{equation}
\begin{split}
\mathcal{L}_\text{CC}&=2G_FV_{ud}\bigg[F_0\partial_\mu\pi^-+i\left(\pi^-\partial_\mu\pi_0-\pi_0\partial_\mu\pi^-\right)+\frac{2}{3F_0}\Big(\pi^-\pi^-\partial_\mu\pi^+-\pi^+\pi^-\partial_\mu\pi^-\\
&\ \ \ + \pi_0\pi^-\partial_\mu\pi_0-\pi^2_0\partial_\mu\pi^-\Big)\bigg]\,\overline{e_L}\,\gamma_\mu\nu_{eL}.
\end{split}
\end{equation}
Again, the sterile neutrino couples with the hadron through the neutrino mixing in Eq.~\eqref{neutmix}. 

We note that the operators in the second-order Lagrangian only provide the corrections to the constants $F_0$ and $M$. Within the accuracy we work, as done in Ref.~\cite{Cirigliano:2017tvr}, one can replace these constants by their physical values to account for these subleading corrections, such as the diagrams involving a pure mesonic loop and the subleading operator insertions. Therefore, we hide the expression of the subleading Lagrangian and, in what follows, make the substitution $F_0\to f_\pi=92.2$ MeV, $M\to m_\pi=138$ MeV.

Moreover, the light sterile neutrino would be integrated out if its momentum approaches or exceeds $\Lambda_\chi$ before matching to ChPT. This means an additional contribution that describes the neutrino exchange in the region less than $1/\Lambda_\chi$, thus calling the hard neutrino exchange contribution. By dissecting the chiral symmetry of the quark operator in the amplitude with two insertions of the weak charge current, as done in Ref.~\cite{Dekens:2020ttz}, one can obtain the pionic operator with two derivatives~\cite{Cirigliano:2017tvr}, which dominates the contributions of hard neutrinos,
\begin{equation}
    \mathcal{L}_\text{hard}=\frac{5}{6}f_\pi^2\kappa(M_i) \,g_\nu^{\pi\pi}(M_i)\partial^\mu\pi^-\partial_\mu\pi^-\,\overline{e_L}e_L^c\,,
\end{equation}
where $g_\nu^{\pi\pi}$ denotes a LEC encoding hard-neutrino exchange and
\begin{equation}
    \kappa(M_i)=\frac{2G_F^2V_{ud}^2\,\mathcal{U}_{ei}^2M_i}{(4\pi f_\pi)^2}.
\end{equation}
By naive dimensional analysis (NDA) the LEC $g_\nu^{\pi\pi}\sim\mathcal{O}(1)$, which can be fixed by experiment and lattice QCD simulation. Because of the non-perturbative effects of QCD, the mass dependence of $g_\nu^{\pi\pi}$ in the type-\uppercase\expandafter{\romannumeral 1} seesaw mechanism is currently unknown. We note the lattice QCD calculation for the mass-dependent $g_\text{LR}^{\pi\pi}$ of the $\pi-\pi$ coupling in the left-right symmetry model in Ref.~\cite{Tuo:2022hft} and expect the prediction of $g_\nu^{\pi\pi}(M_i)$ from the lattice QCD. Besides the Lagrangian in the pion sector, the hard neutrino can also induce chiral operators involving nucleon fields, which are independent of this work.

Similar to the hard-neutrino case, the sterile neutrino with mass $M_i > \Lambda_\chi$ is no longer an appropriate low-energy degree of freedom in ChPT. Integrating out the heavy neutrino in LEFT gives rise to a dimension-nine operator that has the same form as the $O_1$ operator defined in Refs.~\cite{Cirigliano:2017djv, Cirigliano:2018yza}. As a consequence, the heavy-neutrino contribution relevant to this work---parameterized through a LEC $g_1^{\pi\pi}$---can be expressed as in ChPT
\begin{equation}
    \mathcal{L}_\text{heavy}=\frac{5}{6}f_\pi^2\widetilde{\kappa} \,g_1^{\pi\pi}\partial^\mu\pi^-\partial_\mu\pi^-\,\overline{e}_Le_L^c\,,
\end{equation}
where we have ignored the scale dependence of the matching coefficient because this dependence is mild~\cite{Dekens:2023iyc, Dekens:2024hlz}, and
\begin{equation}
    \widetilde{\kappa}=-\frac{4G_F^2V_{ud}^2\,\mathcal{U}_{ei}}{M_i}.
\end{equation}

We are now in a position to calculate the amplitude for the process $\pi^-(p_a)\pi^-(p_b)\to e^-(p_1)e^-(p_2)$ with massless electrons $p_1^2=p_2^2=0$ in the type-\uppercase\expandafter{\romannumeral 1} seesaw mechanism. Similar to the case of the light-neutrino exchange, the amplitude of the transition mediated by sterile neutrinos is straightforward via performing a substitution
\begin{equation}
    \sum_{i=1}^3\frac{U_{ei}^2m_i}{q^2-m_i^2}\longrightarrow\sum_{i=1}^{3+k}\frac{\mathcal{U}_{ei}^2M_i}{q^2-M_i^2}.
\end{equation}
Following the Ref.~\cite{Cirigliano:2017tvr}, we define the transition amplitude as follows
\begin{equation}
    T_{\pi\pi}=2f_\pi^2\sum_{i=1}^{3+k}\,T_\text{lept}(M_i)\,S_{\pi\pi}(M_i),
    \label{totalamp}
\end{equation}
where $T_\text{lept}=4G_F^2V_{ud}^2\,\overline{u}_L(p_1)u_L^c(p_2)\,\mathcal{U}_{ei}^2M_i$ with the spinor of electron $u_L(p)$. The dimensionless amplitude $S_{\pi\pi}$ can be expanded in the powers of momentum
\begin{equation}
    S_{\pi\pi}=S_{\pi\pi}^{(0)}+S_{\pi\pi}^{(2)}+\cdots,
\end{equation}
where $S_{\pi\pi}^{(\nu)}$ denotes the correction with magnitude of $\mathcal{O}(Q^\nu/\Lambda^\nu_\chi)$, called N$^\nu$LO correction in what follows for convenience.

\begin{figure}[htbp]
    \centering
    \begin{subfigure}{0.15\textwidth}
        \centering
        \includegraphics[width=\textwidth]{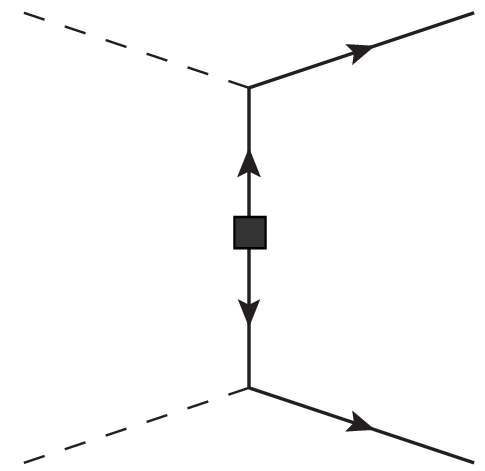}
        \caption{}
    \end{subfigure}
    \begin{subfigure}{0.15\textwidth}
        \centering
        \includegraphics[width=\textwidth]{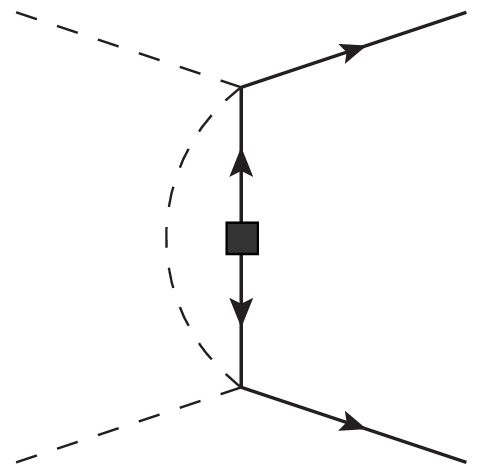}
        \caption{}
    \end{subfigure}
    \begin{subfigure}{0.15\textwidth}
        \centering
        \includegraphics[width=0.9\textwidth]{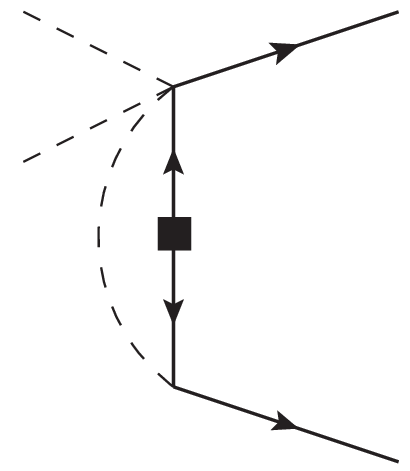}
     \caption{}
    \end{subfigure}
   \begin{subfigure}{0.15\textwidth}
        \centering
        \includegraphics[width=1.1\textwidth]{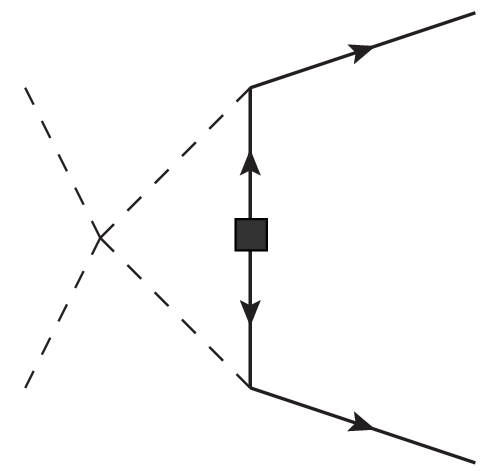}
        \caption{}
    \end{subfigure}
    \begin{subfigure}{0.15\textwidth}
        \centering
        \includegraphics[width=1.7\textwidth]{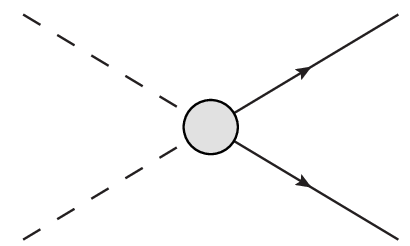}
        \caption{\raggedleft}
    \end{subfigure}
   \caption{\raggedright The diagrams of transition $\pi^-+\pi^-\to e^-+e^-$ up to N$^2$LO. The dashed and solid lines represent pion and lepton fields, respectively. The black square denotes an insertion of the neutrino Majorana mass, and the gray circles sketch the contributions of heavy or hard neutrinos.}
   \label{fig:pipiee}
\end{figure}

For the case of the light sterile neutrino mediating the LNV decay of pion, the diagrams up to N$^2$LO are shown in Fig.\ref{fig:pipiee}.
The Fig.\ref{fig:pipiee}(a) is the tree-level contribution and these remains are subleading contributions. Unlike in the case of nuclear {\onbb} decay, the hard-neutrino contribution occurs at N$^2$LO and cancels the ultraviolet divergence (UV) of the one-loop diagrams. Again, additional one-loop level diagrams extended from Fig.\ref{fig:pipiee}(a), which are not explicitly shown, provide the corrections to the constants $F_0$ and $M$, and these corrections are incorporated by taking their physical values.

Adopting the Mandelstam symbols
\begin{equation}
    s=(p_a+p_b)^2\,,\,\,\,\,t=(p_a-p_1)^2\,,\,\,\,\,u=(p_a-p_2)^2,
\end{equation}
the tree-level amplitude can be written as
\begin{equation}
    S_{\pi\pi}^{(0)}=S_{\pi\pi}^a=-\frac{1}{4}\left(\frac{1}{t-M_i^2}+\frac{1}{u-M_i^2}\right)\left(s-2m_\pi^2\right).
\end{equation}
At N$^2$LO, the amplitude is composed of the one-loop and hard-neutrino contributions:
\begin{equation}
\begin{split}
    S_{\pi\pi}^{(2)}&=S_{\pi\pi}^\text{loop}+S_{\pi\pi}^\text{hard},
\end{split}
\end{equation}
in which the one-loop contribution $S_{\pi\pi}^\text{loop}$ is sum of the following amplitudes
\begin{equation}
    \begin{split}
        &S_{\pi\pi}^b=-\frac{i}{2f_\pi^2}\int\frac{d^4q}{(2\pi)^4}\left[\frac{q^2+q\cdot(p_b-p_a)-p_a\cdot p_b}{(q^2-m_\pi^2)\left[(p_b-p_2-q)^2-M_i^2\right]}+\frac{q^2+q\cdot(p_a-p_b)-p_a\cdot p_b}{(q^2-m_\pi^2)\left[(p_a-p_2-q)^2-M_i^2\right]}\right],\\
        &S_{\pi\pi}^c=\,\frac{i}{3f_\pi^2}\int\frac{d^4q}{(2\pi)^4}\left[\frac{2q^2+q\cdot(p_b+p_a)}{(q^2-m_\pi^2)\left[(p_1-q)^2-M_i^2\right]}+\frac{2q^2-q\cdot(p_a+p_b)}{(q^2-m_\pi^2)\left[(p_2+q)^2-M_i^2\right]}\right],\\
        &S_{\pi\pi}^d=-\frac{i}{3f_\pi^2}\int\frac{d^4q}{(2\pi)^4}\frac{q\cdot(p_b+p_a-q)\left[2p_a\cdot p_b+q\cdot(p_b+p_a-q)\right]}{(q^2-m_\pi^2)\left[(p_1-q)^2-M_i^2\right]\left[(p_a+p_b-q)^2-m_\pi^2\right]}.
    \end{split}
    \label{amps}
\end{equation}
The full expressions of the above amplitudes based on the Mandelstam variables are not listed explicitly because of their complexity. For the hard-neutrino contribution, it can be written as
\begin{equation}
    S_{\pi\pi}^\text{hard}=-\frac{1}{(4\pi f_\pi)^2}\left[\frac{5}{6}g_\nu^{\pi\pi}(M_i,\mu)\frac{s-2m_\pi^2}{2}\right].
\end{equation}
Using the $\overline{\text{MS}}$ scheme, the hard-neutrino term absorbs the UV divergences such that $S_{\pi\pi}^{(2)}$ is a finite and $\mu$-independent quantity. 

At the threshold, corresponding to $s=4m_\pi^2$, $t=u=-m_\pi^2$, the above amplitudes can be written as
\begin{equation}
    \begin{split}
        &S_{\pi\pi}^{(0)}=\frac{m_\pi^2}{m_\pi^2+M_i^2}\\
        &S_{\pi\pi}^{(2)}=-\frac{m_\pi^2}{(4\pi f_\pi)^2}\left[\mathcal{V}_{\pi\pi}(x_i)+\frac{5}{6}\widetilde{g}\,^{\pi\pi}_\nu(M_i)\right]
    \end{split}
\end{equation}
where $\widetilde{g}_\nu^{\pi\pi}(0)=0.45$ obtained by matching on the lattice result in Ref.~\cite{Feng:2018pdq} and $x_i=M_i/m_\pi$, the function $\mathcal{V}_{\pi\pi}(x)$ is defined as 
\begin{widetext}
\begin{equation}
    \begin{split}
        \mathcal{V}_{\pi\pi}(x)=\frac{1}{8}&\left[-24\log(x^2)+2\text{Li}_2\left(\frac{\left(x^2-1\right)^2}{\left(x^2+1\right)^2}\right)-4 \text{Li}_2\left(\frac{x^2-1}{x^2+1}\right)\right.\\
        &\left.-2\text{Li}_2\left(\frac{1}{2} \left(x^2+3\right)\right)+2\text{Li}_2\left(\frac{x^2+3}{x^2+1}\right)-2\text{Li}_2\left(\frac{x^4+2x^2-3}{\left(x^2+1\right)^2}\right)\right.\\
        &\left.+\frac{4 \left(5x^2-7\right)}{x^2-1}-\log ^2\left(\frac{1}{2}\left(-x^2-1\right)\right)-\frac{8 \left(2 x^2-3\right) \log\left(x^2\right)}{\left(x^2-1\right)^2}\right]
    \end{split}
\end{equation}
\end{widetext}
with dilogarithm function $\text{Li}_2(x)$. As $x_i\to0$, the result under the standard mechanism given by Ref.~\cite{Cirigliano:2017tvr} will be recovered.

One of the main features of the seesaw model is that the ChPT description originating from this model is based not only on the pre-set expansion in $Q/\Lambda_\chi$ but also on another expansion in $M_i/\Lambda_\chi$. This is because the loop integral gives rise to the positive power of the neutrino mass such that this additional expansion becomes possible. For the light neutrino, $M_i/\Lambda_\chi$ is an expansion parameter of ChPT. Therefore, one can expand the LEC $\widetilde{g}_\nu^{\pi\pi}$ in the powers of $M_i$ as follows~\cite{Dekens:2024hlz}
\begin{equation}
    \widetilde{g}_\nu^{\pi\pi}(M_i)=\widetilde{g}_\nu^{\pi\pi}(0)\left[1+\mathcal{O}(M_i^2/\Lambda_\chi^2)\right].
    \label{lecerr}
\end{equation}
The second term in the above expansion can be regarded as a low-mass truncation error of ChPT. Moreover, by NDA, the amplitude up to the one-loop level we calculate accepts an additional uncertainty with $\mathcal{O}(M_i^4/\Lambda^4_\chi)$ arising from the next order such as two-loop corrections. As a result, the amplitude with error can be expressed as
\begin{equation}
    S_{\pi\pi}=\left\{S_{\pi\pi}^{(0)}(M_i)+S_{\pi\pi}^\text{loop}(M_i)+S_{\pi\pi}^\text{hard}(0)\left[1+\mathcal{O}\left(\frac{M_i^2}{\Lambda_\chi^2}\right)\right]\right\}\left[1+\mathcal{O}\left(\frac{M_i^4}{\Lambda_\chi^4}\right)\right].
    \label{ampuncert}
\end{equation}
It should be emphasized that the procedure above adopts a conservative way to estimate the effects of the mass of sterile neutrino on the amplitude. The construction of a consistent power counting that accommodates the massive sterile neutrino effects remains a crucial object in ChPT.

In the context of the heavy neutrino, the amplitude of the pion decay up to N$^2$LO is given by
\begin{equation}
    S_{\pi\pi}^\text{heavy}=\frac{1}{M_i^2}\left[\frac{5}{6}g_1^{\pi\pi}\left(s-2m_\pi^2\right)\right],
\end{equation}
where $g_1^{\pi\pi}=0.36$ at the scale $\mu=2$ GeV extracted from the lattice calculation in Ref.~\cite{Nicholson:2018mwc}.

\subsection{Interpolation Formula}

In principle, one now obtains the mass-dependent amplitude of the pionic {\onbb} decay within the type-\uppercase\expandafter{\romannumeral 1} seesaw mechanism based on ChPT. However, due to the expansion parameter $M_i/\Lambda_\chi$, the ChPT description would break down when the neutrino mass approaches $\Lambda_\chi$. Consequently, the amplitude of the pion decay in the region where the neutrino mass is close to $\Lambda_\chi$ needs to be modeled.

There is an intuition that the evolution of the amplitude from light-neutrino to heavy-neutrino area is continuous. Therefore, with the constraint of the heavy-neutrino asymptotic behavior
\begin{equation}
    \lim_{M_i\to\infty}S_{\pi\pi}(M_i)=S_{\pi\pi}^\text{heavy}\propto\frac{m_\pi^2}{M_i^2},
\end{equation}
one can construct the following $M_i$ dependence in the $M_i<\Lambda_\chi$ region 
\begin{equation}
    S_{\pi\pi}^\text{fit}(M_i)=S_{\pi\pi}(0)\frac{1}{1+(M_i/m_a)^2}
    \label{modslong}
\end{equation}
where $S_{\pi\pi}(0)=S^{(0)}_{\pi\pi}(0)+S^{(2)}_{\pi\pi}(0)$ given by ChPT. Based on the above intuition, one can utilize the matching condition
\begin{equation}
    S_{\pi\pi}^\text{fit}(M_i)=S_{\pi\pi}^\text{heavy}(M_i)
    \label{lhmc}
\end{equation}
to pin down the parameter $m_a$ at $M_i=2$ GeV. 

The reason for considering Eq.\eqref{modslong} is that this interpolation formula, which has been widely used to predict the mass-dependent NMEs of the nuclear {\onbb} decay, e.g. see Refs~\cite{Faessler:2014kka, Barea:2015zfa, Asaka:2016zib, Babic:2018ikc, Asaka:2020wfo, Fang:2021jfv, Fang:2024hzy}, captures the main features of the amplitude in the light- and heavy-neutrino region. Because of the unknown dependence of the LEC $g_\nu^{\pi\pi}$ on the neutrino mass so far, we do not plan to discern the $M_i$ dependence of the long- and short-range contributions as done in Refs.~\cite{Dekens:2020ttz, Dekens:2023iyc, Dekens:2024hlz}.

\section{Results and discussions}\label{sec:resluts}
The actual breakdown scale of EFT is often lower than the theoretical expectation $\Lambda_\chi$ for a given process. For example, in $\pi - \pi$ scattering, the theoretical range is associated with the $\rho$-meson mass, while in nucleon-nucleon scattering this range is approximately 500 MeV~\cite{Mehen:1998tp}. Determining the realistic breakdown scale of ChPT in a specific process is essential because it limits the range of theory and provides an input to the uncertainty estimation. To this end, we first focus on the mass dependence of the tree-level and one-loop contributions to the process in which two pions convert to two electrons at the threshold under the light sterile neutrino exchange mechanism.

\begin{figure}[htbp]
    \centering
    \includegraphics[scale=0.55]{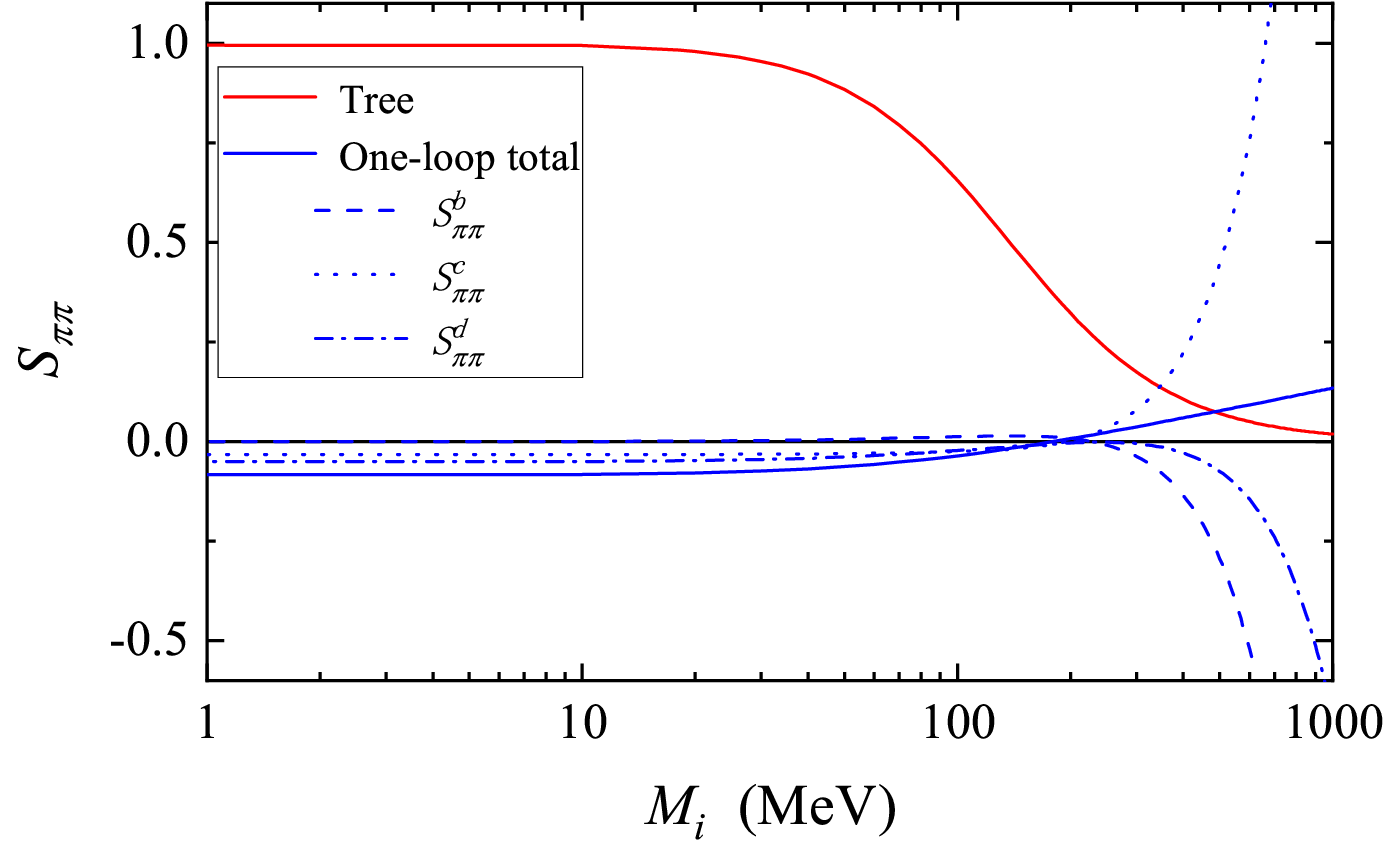}
    \caption{\raggedright Dependence of the threshold amplitude $S_{\pi\pi}$ on the sterile neutrino mass $M_i$, showing the contributions from tree-level and one-loop diagrams (sub-diagrams (b), (c), and (d)) based on ChPT calculations.}
    \label{fig:lovs}
\end{figure}

As shown in Fig.\ref{fig:lovs}, both tree-level and one-loop contributions in $M_i\lesssim 50$ MeV exhibit less sensitivity to neutrino mass compared to the regime of $M_i\sim m_\pi$. This is unsurprising because the neutrino mass is smaller than the typical momentum $\sim m_\pi$ in these cases.
Furthermore, the one-loop contribution becomes zero for $M_i\simeq 200$ MeV $\sim m_\pi$. In the range of $M_i\gtrsim m_\pi$, the tree-level amplitude still decreases with increasing $M_i$, whereas the one-loop correction increases logarithmically. When $M_i$ approaches $500\ \text{MeV}$, we find the magnitude of the one-loop correction is comparable to the tree-level contribution. In principle, up to this scale, the mesons with masses heavier than $m_\pi$ (such as K and $\eta$ mesons) can be created and should be included as the explicit degrees of freedom in ChPT. One naturally expects that the one-loop correction may be changed due to the contributions of these heavier mesons, thereby ensuring the validity of the perturbative expansion. However, by means of the chiral Lagrangian in the three-flavor case, we confirmed up to the one-loop level that the K and $\eta$ mesons do not generate extra contributions to this process.

To gain insight into the effects of various one-loop diagrams on this sub-leading contribution, we illustrate in Fig.~\ref{fig:lovs} the mass dependence of the different amplitudes $S_{\pi\pi}^b$, $S_{\pi\pi}^c$ and $S_{\pi\pi}^d$ in eq.\eqref{amps} denoted by dotted, dashed, and dot-dash lines, respectively. The mass independence of these contributions from three diagrams can be observed in the range of $M_i\lesssim 50$ MeV. Meanwhile, $S_{\pi\pi}^b$ is nearly negligible, and $S_{\pi\pi}^c$ is almost equivalent to the contribution of $S_{\pi\pi}^d$. Notably, for $M_i\simeq 200$ MeV $\sim m_\pi$, the contributions of three one-loop diagrams approach zero, leading to a tiny one-loop correction at this point. This behavior arises because a loop integral in eq.\eqref{amps} vanishes after renormalization when $M_i$ approaches $m_\pi$. In the heavier mass region, the contributions of various diagrams change dramatically, and the cancellation between the $S_{\pi\pi}^c$ and the remaining amplitudes leads to a mild mass dependence of the total correction (a logarithmical increase with $M_i$). Moreover, $S_{\pi\pi}^b$ and $S_{\pi\pi}^c$ will exceed the tree-level contribution when $M_i\gtrsim 400$ MeV. This simply implies that the neutrino with mass $M_i\gtrsim 3m_\pi$ is no longer a suitable light degree of freedom in ChPT and it may be possible to treat the neutrino as an explicit heavy degree of freedom in a way similar to the treatment of the $\Delta(1232)$ resonance in ChPT.

Apart from the one-loop correction at N$^2$LO, there is a contact contribution that stems from the hard-neutrino exchange. According to the interpolation formulae of the hard-neutrino LECs discussed in Ref.~\cite{Dekens:2020ttz} and the lattice simulation~\cite{Tuo:2022hft}, one can foresee that the hard-neutrino contribution is negatively related to $M_i$. On the other hand, this contact term doesn't change the situation of the breaking of perturbative expansion for ChPT. Consequently, we suggest that the breakdown scale $\Lambda$ in this framework is about 500 MeV and $\Lambda_\chi$ will be replaced by $\Lambda$ in the following analysis. 

\begin{figure}[htbp]
    \centering
    \includegraphics[scale=0.55]{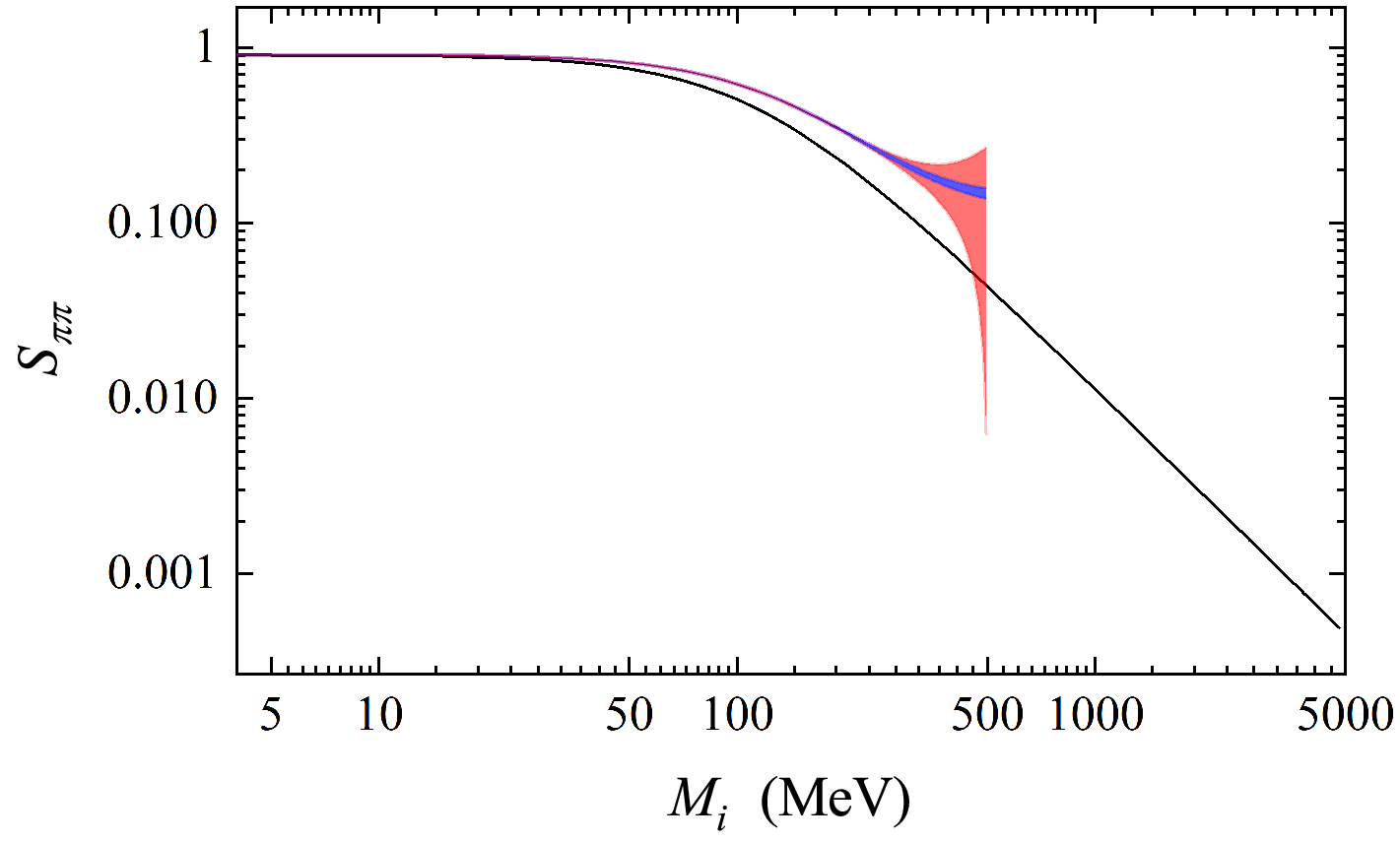}
    \caption{\raggedright The amplitude $S_{\pi\pi}$ at threshold predicted by Eq.\eqref{modslong} as a function of the sterile neutrino mass, in comparison with the amplitude based on ChPT. The blue and red bands represent the results with the low-mass truncation errors of $\widetilde{g}_\nu^{\pi\pi}$ and amplitude, respectively. The black curve denotes the prediction of the interpolation formula.}
    \label{fig:spp}
\end{figure}

We fix the value of the parameter $m_a$ in Eq.~\eqref{modslong} to be 112 MeV from our calculation for the amplitude at the threshold, it is then comparable to the pion mass. In Fig.\ref{fig:spp}, we depict the $M_i$ dependence of the amplitude $S_{\pi\pi}$ based on the ChPT and Eq.\eqref{modslong}, respectively. 

To survey the validity of the interpolation formula, the uncertainty estimation based on Eq.\eqref{ampuncert} is also included, in which the blue (red) band denotes the low-mass truncation error of $\widetilde{g}_\nu^{\pi\pi}$ (amplitude). Obviously, the truncation error of $\widetilde{g}_\nu^{\pi\pi}$ given by Eq.\eqref{lecerr} has a minor effect on the amplitude, this again suggests that the contribution of the hard neutrino is hard to cure the breakdown of the expansion in $M_i$. Moreover, the ChPT result, as specified by Eq.\eqref{ampuncert}, remains nearly invariant in the $M_i \lesssim 50\ \text{MeV}$ region, and the interpolation formula is in excellent agreement with the ChPT calculations. However, the difference between this interpolation and ChPT predictions emerges when $M_i\gtrsim50$ MeV. Especially for the mass region where ChPT is still valid such as $M_i\simeq m_\pi$, the prediction of the interpolation formula deviates significantly from the ChPT results, meaning a demand for a more reliable interpolation formula. Furthermore, the low-mass truncation error of the amplitude displays a divergent trend in the region with $M_i\gtrsim300$ MeV, reflecting the conclusions drawn from Fig. \ref{fig:lovs} regarding the breakdown of the perturbative approach. For the heavy neutrino, the amplitude $S_{\pi\pi}$ from the interpolation is consistent with ChPT.

As mentioned in Sec.\ref{sec:theory}, the interpolation formula Eq.\eqref{modslong} uses only the data of two extreme cases, the amplitudes under the light- and heavy-neutrino exchange, to predict the mass dependence of the amplitude. Although correctly describing the main behavior of the amplitude, this method may be under-fitting. In addition, Eq.\eqref{modslong} adopts the most economic assumption that the ratio of the long-range contribution over the hard-neutrino contribution is independent of $M_i$. This means in the case of the light sterile neutrino that $g_\nu^{\pi\pi}$ can be expressed as an expansion in powers of $M_i/m_a$ rather than $M_i/\Lambda$, which contradicts NDA. Therefore, the form of Eq.\eqref{modslong} is insufficient for the pionic {\onbb} decay mediated by a sterile neutrino. We also note that a similar conclusion is shown in the two-nucleon case by Refs.~\cite{Dekens:2023iyc, Dekens:2024hlz}.

\section{Summary}\label{sec:summary}

In this work, we studied the pionic {\onbb} decay induced by the sterile neutrinos with masses $M_i$ in the framework of ChPT. The sterile neutrino mass has not been stringently limited in experimental and theoretical aspects so far such that we explore this process in a broad mass range. Since the mass dependence of the amplitude can be derived analytically in the case of the heavy sterile neutrino, we focused on the mass region below a few GeV. We calculated the transition amplitude under the light- and heavy-neutrino mechanism up to N$^2$LO.

The amplitude associated with heavy neutrinos up to N$^2$LO is given by the contact term. Its mass dependence is scaled as $M_i^{-2}$. For the scenario of light neutrino, which is an active degree of freedom in ChPT, the one-loop correction as a subleading contribution would be over the leading-order contribution in the intermediate mass region with $500\,\text{MeV} \lesssim M_i\lesssim\Lambda_\chi$. Thus, we recommend $\Lambda=500$ MeV as the breaking scale of ChPT in the LNV pion decay process mediated by light sterile neutrino. 

Because of the ChPT expansion breakdown in the intermediate mass region, we employed a naive interpolation formula to connect the amplitude based on ChPT across the light and heavy sterile neutrino mass regions. This interpolation formula, which assumes the long- and short-range contributions share the same functional form of mass, does not recover the ChPT prediction well in the light mass region. Differing from it, a sort of interpolation that distinguishes the mass dependence of the long- and short-range contributions is presented in Refs.~\cite{Dekens:2020ttz, Dekens:2023iyc, Dekens:2024hlz}. More studies on the mass dependence of the amplitude, especially for short-range contributions, are needed to describe accurately the {\onbb} decay under the seesaw mechanism.

\begin{acknowledgments}

This work is supported by National Key Research and Development Program of China (2021YFA1601300), Young Scientists in Basic Research (YSBR-099) from Chinese Academy of Sciences and the Postdoctoral Special Foundation of Gansu Province.

\end{acknowledgments}

\bibliography{piondecay}

\begin{thebibliography}{46}%
\makeatletter
\providecommand \@ifxundefined [1]{%
 \@ifx{#1\undefined}
}%
\providecommand \@ifnum [1]{%
 \ifnum #1\expandafter \@firstoftwo
 \else \expandafter \@secondoftwo
 \fi
}%
\providecommand \@ifx [1]{%
 \ifx #1\expandafter \@firstoftwo
 \else \expandafter \@secondoftwo
 \fi
}%
\providecommand \natexlab [1]{#1}%
\providecommand \enquote  [1]{``#1''}%
\providecommand \bibnamefont  [1]{#1}%
\providecommand \bibfnamefont [1]{#1}%
\providecommand \citenamefont [1]{#1}%
\providecommand \href@noop [0]{\@secondoftwo}%
\providecommand \href [0]{\begingroup \@sanitize@url \@href}%
\providecommand \@href[1]{\@@startlink{#1}\@@href}%
\providecommand \@@href[1]{\endgroup#1\@@endlink}%
\providecommand \@sanitize@url [0]{\catcode `\\12\catcode `\$12\catcode
  `\&12\catcode `\#12\catcode `\^12\catcode `\_12\catcode `\%12\relax}%
\providecommand \@@startlink[1]{}%
\providecommand \@@endlink[0]{}%
\providecommand \url  [0]{\begingroup\@sanitize@url \@url }%
\providecommand \@url [1]{\endgroup\@href {#1}{\urlprefix }}%
\providecommand \urlprefix  [0]{URL }%
\providecommand \Eprint [0]{\href }%
\providecommand \doibase [0]{http://dx.doi.org/}%
\providecommand \selectlanguage [0]{\@gobble}%
\providecommand \bibinfo  [0]{\@secondoftwo}%
\providecommand \bibfield  [0]{\@secondoftwo}%
\providecommand \translation [1]{[#1]}%
\providecommand \BibitemOpen [0]{}%
\providecommand \bibitemStop [0]{}%
\providecommand \bibitemNoStop [0]{.\EOS\space}%
\providecommand \EOS [0]{\spacefactor3000\relax}%
\providecommand \BibitemShut  [1]{\csname bibitem#1\endcsname}%
\let\auto@bib@innerbib\@empty
\bibitem [{\citenamefont {Fukuda}\ \emph {et~al.}(1998)\citenamefont {Fukuda}
  \emph {et~al.}}]{Super-Kamiokande:1998kpq}%
  \BibitemOpen
  \bibfield  {author} {\bibinfo {author} {\bibfnamefont {Y.}~\bibnamefont
  {Fukuda}} \emph {et~al.} (\bibinfo {collaboration} {Super-Kamiokande}),\
  }\href {\doibase 10.1103/PhysRevLett.81.1562} {\bibfield  {journal} {\bibinfo
   {journal} {Phys. Rev. Lett.}\ }\textbf {\bibinfo {volume} {81}},\ \bibinfo
  {pages} {1562} (\bibinfo {year} {1998})},\ \Eprint
  {http://arxiv.org/abs/hep-ex/9807003} {arXiv:hep-ex/9807003} \BibitemShut
  {NoStop}%
\bibitem [{\citenamefont {Ahmad}\ \emph {et~al.}(2001)\citenamefont {Ahmad}
  \emph {et~al.}}]{SNO:2001kpb}%
  \BibitemOpen
  \bibfield  {author} {\bibinfo {author} {\bibfnamefont {Q.~R.}\ \bibnamefont
  {Ahmad}} \emph {et~al.} (\bibinfo {collaboration} {SNO}),\ }\href {\doibase
  10.1103/PhysRevLett.87.071301} {\bibfield  {journal} {\bibinfo  {journal}
  {Phys. Rev. Lett.}\ }\textbf {\bibinfo {volume} {87}},\ \bibinfo {pages}
  {071301} (\bibinfo {year} {2001})},\ \Eprint
  {http://arxiv.org/abs/nucl-ex/0106015} {arXiv:nucl-ex/0106015} \BibitemShut
  {NoStop}%
\bibitem [{\citenamefont {Ahn}\ \emph {et~al.}(2003)\citenamefont {Ahn} \emph
  {et~al.}}]{K2K:2002icj}%
  \BibitemOpen
  \bibfield  {author} {\bibinfo {author} {\bibfnamefont {M.~H.}\ \bibnamefont
  {Ahn}} \emph {et~al.} (\bibinfo {collaboration} {K2K}),\ }\href {\doibase
  10.1103/PhysRevLett.90.041801} {\bibfield  {journal} {\bibinfo  {journal}
  {Phys. Rev. Lett.}\ }\textbf {\bibinfo {volume} {90}},\ \bibinfo {pages}
  {041801} (\bibinfo {year} {2003})},\ \Eprint
  {http://arxiv.org/abs/hep-ex/0212007} {arXiv:hep-ex/0212007} \BibitemShut
  {NoStop}%
\bibitem [{\citenamefont {Minkowski}(1977)}]{Minkowski:1977sc}%
  \BibitemOpen
  \bibfield  {author} {\bibinfo {author} {\bibfnamefont {P.}~\bibnamefont
  {Minkowski}},\ }\href {\doibase 10.1016/0370-2693(77)90435-X} {\bibfield
  {journal} {\bibinfo  {journal} {Phys. Lett. B}\ }\textbf {\bibinfo {volume}
  {67}},\ \bibinfo {pages} {421} (\bibinfo {year} {1977})}\BibitemShut
  {NoStop}%
\bibitem [{\citenamefont {Yanagida}(1979)}]{Yanagida:1979as}%
  \BibitemOpen
  \bibfield  {author} {\bibinfo {author} {\bibfnamefont {T.}~\bibnamefont
  {Yanagida}},\ }\href@noop {} {\bibfield  {journal} {\bibinfo  {journal}
  {Conf. Proc. C}\ }\textbf {\bibinfo {volume} {7902131}},\ \bibinfo {pages}
  {95} (\bibinfo {year} {1979})}\BibitemShut {NoStop}%
\bibitem [{\citenamefont {Gell-Mann}\ \emph {et~al.}(1979)\citenamefont
  {Gell-Mann}, \citenamefont {Ramond},\ and\ \citenamefont
  {Slansky}}]{Gell-Mann:1979vob}%
  \BibitemOpen
  \bibfield  {author} {\bibinfo {author} {\bibfnamefont {M.}~\bibnamefont
  {Gell-Mann}}, \bibinfo {author} {\bibfnamefont {P.}~\bibnamefont {Ramond}}, \
  and\ \bibinfo {author} {\bibfnamefont {R.}~\bibnamefont {Slansky}},\
  }\href@noop {} {\bibfield  {journal} {\bibinfo  {journal} {Conf. Proc. C}\
  }\textbf {\bibinfo {volume} {790927}},\ \bibinfo {pages} {315} (\bibinfo
  {year} {1979})},\ \Eprint {http://arxiv.org/abs/1306.4669} {arXiv:1306.4669
  [hep-th]} \BibitemShut {NoStop}%
\bibitem [{\citenamefont {Glashow}(1980)}]{Glashow:1979nm}%
  \BibitemOpen
  \bibfield  {author} {\bibinfo {author} {\bibfnamefont {S.~L.}\ \bibnamefont
  {Glashow}},\ }\href {\doibase 10.1007/978-1-4684-7197-7_15} {\bibfield
  {journal} {\bibinfo  {journal} {NATO Sci. Ser. B}\ }\textbf {\bibinfo
  {volume} {61}},\ \bibinfo {pages} {687} (\bibinfo {year} {1980})}\BibitemShut
  {NoStop}%
\bibitem [{\citenamefont {Mohapatra}\ and\ \citenamefont
  {Senjanovic}(1980)}]{Mohapatra:1979ia}%
  \BibitemOpen
  \bibfield  {author} {\bibinfo {author} {\bibfnamefont {R.~N.}\ \bibnamefont
  {Mohapatra}}\ and\ \bibinfo {author} {\bibfnamefont {G.}~\bibnamefont
  {Senjanovic}},\ }\href {\doibase 10.1103/PhysRevLett.44.912} {\bibfield
  {journal} {\bibinfo  {journal} {Phys. Rev. Lett.}\ }\textbf {\bibinfo
  {volume} {44}},\ \bibinfo {pages} {912} (\bibinfo {year} {1980})}\BibitemShut
  {NoStop}%
\bibitem [{\citenamefont {Asaka}\ and\ \citenamefont
  {Shaposhnikov}(2005)}]{Asaka:2005pn}%
  \BibitemOpen
  \bibfield  {author} {\bibinfo {author} {\bibfnamefont {T.}~\bibnamefont
  {Asaka}}\ and\ \bibinfo {author} {\bibfnamefont {M.}~\bibnamefont
  {Shaposhnikov}},\ }\href {\doibase 10.1016/j.physletb.2005.06.020} {\bibfield
   {journal} {\bibinfo  {journal} {Phys. Lett. B}\ }\textbf {\bibinfo {volume}
  {620}},\ \bibinfo {pages} {17} (\bibinfo {year} {2005})},\ \Eprint
  {http://arxiv.org/abs/hep-ph/0505013} {arXiv:hep-ph/0505013} \BibitemShut
  {NoStop}%
\bibitem [{\citenamefont {Shaposhnikov}(2008)}]{Shaposhnikov:2008pf}%
  \BibitemOpen
  \bibfield  {author} {\bibinfo {author} {\bibfnamefont {M.}~\bibnamefont
  {Shaposhnikov}},\ }\href {\doibase 10.1088/1126-6708/2008/08/008} {\bibfield
  {journal} {\bibinfo  {journal} {JHEP}\ }\textbf {\bibinfo {volume} {08}},\
  \bibinfo {pages} {008} (\bibinfo {year} {2008})},\ \Eprint
  {http://arxiv.org/abs/0804.4542} {arXiv:0804.4542 [hep-ph]} \BibitemShut
  {NoStop}%
\bibitem [{\citenamefont {Schechter}\ and\ \citenamefont
  {Valle}(1982)}]{Schechter:1981bd}%
  \BibitemOpen
  \bibfield  {author} {\bibinfo {author} {\bibfnamefont {J.}~\bibnamefont
  {Schechter}}\ and\ \bibinfo {author} {\bibfnamefont {J.~W.~F.}\ \bibnamefont
  {Valle}},\ }\href {\doibase 10.1103/PhysRevD.25.2951} {\bibfield  {journal}
  {\bibinfo  {journal} {Phys. Rev. D}\ }\textbf {\bibinfo {volume} {25}},\
  \bibinfo {pages} {2951} (\bibinfo {year} {1982})}\BibitemShut {NoStop}%
\bibitem [{\citenamefont {Agostini}\ \emph {et~al.}(2020)\citenamefont
  {Agostini} \emph {et~al.}}]{GERDA:2020xhi}%
  \BibitemOpen
  \bibfield  {author} {\bibinfo {author} {\bibfnamefont {M.}~\bibnamefont
  {Agostini}} \emph {et~al.} (\bibinfo {collaboration} {GERDA}),\ }\href
  {\doibase 10.1103/PhysRevLett.125.252502} {\bibfield  {journal} {\bibinfo
  {journal} {Phys. Rev. Lett.}\ }\textbf {\bibinfo {volume} {125}},\ \bibinfo
  {pages} {252502} (\bibinfo {year} {2020})},\ \Eprint
  {http://arxiv.org/abs/2009.06079} {arXiv:2009.06079 [nucl-ex]} \BibitemShut
  {NoStop}%
\bibitem [{\citenamefont {Abe}\ \emph {et~al.}(2023)\citenamefont {Abe} \emph
  {et~al.}}]{KamLAND-Zen:2022tow}%
  \BibitemOpen
  \bibfield  {author} {\bibinfo {author} {\bibfnamefont {S.}~\bibnamefont
  {Abe}} \emph {et~al.} (\bibinfo {collaboration} {KamLAND-Zen}),\ }\href
  {\doibase 10.1103/PhysRevLett.130.051801} {\bibfield  {journal} {\bibinfo
  {journal} {Phys. Rev. Lett.}\ }\textbf {\bibinfo {volume} {130}},\ \bibinfo
  {pages} {051801} (\bibinfo {year} {2023})},\ \Eprint
  {http://arxiv.org/abs/2203.02139} {arXiv:2203.02139 [hep-ex]} \BibitemShut
  {NoStop}%
\bibitem [{\citenamefont {Abgrall}\ \emph {et~al.}(2017)\citenamefont {Abgrall}
  \emph {et~al.}}]{LEGEND:2017cdu}%
  \BibitemOpen
  \bibfield  {author} {\bibinfo {author} {\bibfnamefont {N.}~\bibnamefont
  {Abgrall}} \emph {et~al.} (\bibinfo {collaboration} {LEGEND}),\ }\href
  {\doibase 10.1063/1.5007652} {\bibfield  {journal} {\bibinfo  {journal} {AIP
  Conf. Proc.}\ }\textbf {\bibinfo {volume} {1894}},\ \bibinfo {pages} {020027}
  (\bibinfo {year} {2017})},\ \Eprint {http://arxiv.org/abs/1709.01980}
  {arXiv:1709.01980 [physics.ins-det]} \BibitemShut {NoStop}%
\bibitem [{\citenamefont {Albert}\ \emph {et~al.}(2018)\citenamefont {Albert}
  \emph {et~al.}}]{nEXO:2017nam}%
  \BibitemOpen
  \bibfield  {author} {\bibinfo {author} {\bibfnamefont {J.~B.}\ \bibnamefont
  {Albert}} \emph {et~al.} (\bibinfo {collaboration} {nEXO}),\ }\href {\doibase
  10.1103/PhysRevC.97.065503} {\bibfield  {journal} {\bibinfo  {journal} {Phys.
  Rev. C}\ }\textbf {\bibinfo {volume} {97}},\ \bibinfo {pages} {065503}
  (\bibinfo {year} {2018})},\ \Eprint {http://arxiv.org/abs/1710.05075}
  {arXiv:1710.05075 [nucl-ex]} \BibitemShut {NoStop}%
\bibitem [{\citenamefont {Cirigliano}\ \emph {et~al.}(2017)\citenamefont
  {Cirigliano}, \citenamefont {Dekens}, \citenamefont {de~Vries}, \citenamefont
  {Graesser},\ and\ \citenamefont {Mereghetti}}]{Cirigliano:2017djv}%
  \BibitemOpen
  \bibfield  {author} {\bibinfo {author} {\bibfnamefont {V.}~\bibnamefont
  {Cirigliano}}, \bibinfo {author} {\bibfnamefont {W.}~\bibnamefont {Dekens}},
  \bibinfo {author} {\bibfnamefont {J.}~\bibnamefont {de~Vries}}, \bibinfo
  {author} {\bibfnamefont {M.~L.}\ \bibnamefont {Graesser}}, \ and\ \bibinfo
  {author} {\bibfnamefont {E.}~\bibnamefont {Mereghetti}},\ }\href {\doibase
  10.1007/JHEP12(2017)082} {\bibfield  {journal} {\bibinfo  {journal} {JHEP}\
  }\textbf {\bibinfo {volume} {12}},\ \bibinfo {pages} {082} (\bibinfo {year}
  {2017})},\ \Eprint {http://arxiv.org/abs/1708.09390} {arXiv:1708.09390
  [hep-ph]} \BibitemShut {NoStop}%
\bibitem [{\citenamefont {Cirigliano}\ \emph
  {et~al.}(2018{\natexlab{a}})\citenamefont {Cirigliano}, \citenamefont
  {Dekens}, \citenamefont {de~Vries}, \citenamefont {Graesser},\ and\
  \citenamefont {Mereghetti}}]{Cirigliano:2018yza}%
  \BibitemOpen
  \bibfield  {author} {\bibinfo {author} {\bibfnamefont {V.}~\bibnamefont
  {Cirigliano}}, \bibinfo {author} {\bibfnamefont {W.}~\bibnamefont {Dekens}},
  \bibinfo {author} {\bibfnamefont {J.}~\bibnamefont {de~Vries}}, \bibinfo
  {author} {\bibfnamefont {M.~L.}\ \bibnamefont {Graesser}}, \ and\ \bibinfo
  {author} {\bibfnamefont {E.}~\bibnamefont {Mereghetti}},\ }\href {\doibase
  10.1007/JHEP12(2018)097} {\bibfield  {journal} {\bibinfo  {journal} {JHEP}\
  }\textbf {\bibinfo {volume} {12}},\ \bibinfo {pages} {097} (\bibinfo {year}
  {2018}{\natexlab{a}})},\ \Eprint {http://arxiv.org/abs/1806.02780}
  {arXiv:1806.02780 [hep-ph]} \BibitemShut {NoStop}%
\bibitem [{\citenamefont {Yao}\ \emph {et~al.}(2020)\citenamefont {Yao},
  \citenamefont {Bally}, \citenamefont {Engel}, \citenamefont {Wirth},
  \citenamefont {Rodr\'\i{}guez},\ and\ \citenamefont {Hergert}}]{Yao:2019rck}%
  \BibitemOpen
  \bibfield  {author} {\bibinfo {author} {\bibfnamefont {J.~M.}\ \bibnamefont
  {Yao}}, \bibinfo {author} {\bibfnamefont {B.}~\bibnamefont {Bally}}, \bibinfo
  {author} {\bibfnamefont {J.}~\bibnamefont {Engel}}, \bibinfo {author}
  {\bibfnamefont {R.}~\bibnamefont {Wirth}}, \bibinfo {author} {\bibfnamefont
  {T.~R.}\ \bibnamefont {Rodr\'\i{}guez}}, \ and\ \bibinfo {author}
  {\bibfnamefont {H.}~\bibnamefont {Hergert}},\ }\href {\doibase
  10.1103/PhysRevLett.124.232501} {\bibfield  {journal} {\bibinfo  {journal}
  {Phys. Rev. Lett.}\ }\textbf {\bibinfo {volume} {124}},\ \bibinfo {pages}
  {232501} (\bibinfo {year} {2020})},\ \Eprint
  {http://arxiv.org/abs/1908.05424} {arXiv:1908.05424 [nucl-th]} \BibitemShut
  {NoStop}%
\bibitem [{\citenamefont {Belley}\ \emph {et~al.}(2021)\citenamefont {Belley},
  \citenamefont {Payne}, \citenamefont {Stroberg}, \citenamefont {Miyagi},\
  and\ \citenamefont {Holt}}]{Belley:2020ejd}%
  \BibitemOpen
  \bibfield  {author} {\bibinfo {author} {\bibfnamefont {A.}~\bibnamefont
  {Belley}}, \bibinfo {author} {\bibfnamefont {C.~G.}\ \bibnamefont {Payne}},
  \bibinfo {author} {\bibfnamefont {S.~R.}\ \bibnamefont {Stroberg}}, \bibinfo
  {author} {\bibfnamefont {T.}~\bibnamefont {Miyagi}}, \ and\ \bibinfo {author}
  {\bibfnamefont {J.~D.}\ \bibnamefont {Holt}},\ }\href {\doibase
  10.1103/PhysRevLett.126.042502} {\bibfield  {journal} {\bibinfo  {journal}
  {Phys. Rev. Lett.}\ }\textbf {\bibinfo {volume} {126}},\ \bibinfo {pages}
  {042502} (\bibinfo {year} {2021})},\ \Eprint
  {http://arxiv.org/abs/2008.06588} {arXiv:2008.06588 [nucl-th]} \BibitemShut
  {NoStop}%
\bibitem [{\citenamefont {Novario}\ \emph {et~al.}(2021)\citenamefont
  {Novario}, \citenamefont {Gysbers}, \citenamefont {Engel}, \citenamefont
  {Hagen}, \citenamefont {Jansen}, \citenamefont {Morris}, \citenamefont
  {Navr\'atil}, \citenamefont {Papenbrock},\ and\ \citenamefont
  {Quaglioni}}]{Novario:2020dmr}%
  \BibitemOpen
  \bibfield  {author} {\bibinfo {author} {\bibfnamefont {S.}~\bibnamefont
  {Novario}}, \bibinfo {author} {\bibfnamefont {P.}~\bibnamefont {Gysbers}},
  \bibinfo {author} {\bibfnamefont {J.}~\bibnamefont {Engel}}, \bibinfo
  {author} {\bibfnamefont {G.}~\bibnamefont {Hagen}}, \bibinfo {author}
  {\bibfnamefont {G.~R.}\ \bibnamefont {Jansen}}, \bibinfo {author}
  {\bibfnamefont {T.~D.}\ \bibnamefont {Morris}}, \bibinfo {author}
  {\bibfnamefont {P.}~\bibnamefont {Navr\'atil}}, \bibinfo {author}
  {\bibfnamefont {T.}~\bibnamefont {Papenbrock}}, \ and\ \bibinfo {author}
  {\bibfnamefont {S.}~\bibnamefont {Quaglioni}},\ }\href {\doibase
  10.1103/PhysRevLett.126.182502} {\bibfield  {journal} {\bibinfo  {journal}
  {Phys. Rev. Lett.}\ }\textbf {\bibinfo {volume} {126}},\ \bibinfo {pages}
  {182502} (\bibinfo {year} {2021})},\ \Eprint
  {http://arxiv.org/abs/2008.09696} {arXiv:2008.09696 [nucl-th]} \BibitemShut
  {NoStop}%
\bibitem [{\citenamefont {Belley}\ \emph {et~al.}(2024)\citenamefont {Belley}
  \emph {et~al.}}]{Belley:2023lec}%
  \BibitemOpen
  \bibfield  {author} {\bibinfo {author} {\bibfnamefont {A.}~\bibnamefont
  {Belley}} \emph {et~al.},\ }\href {\doibase 10.1103/PhysRevLett.132.182502}
  {\bibfield  {journal} {\bibinfo  {journal} {Phys. Rev. Lett.}\ }\textbf
  {\bibinfo {volume} {132}},\ \bibinfo {pages} {182502} (\bibinfo {year}
  {2024})},\ \Eprint {http://arxiv.org/abs/2308.15634} {arXiv:2308.15634
  [nucl-th]} \BibitemShut {NoStop}%
\bibitem [{\citenamefont {Agostini}\ \emph {et~al.}(2023)\citenamefont
  {Agostini}, \citenamefont {Benato}, \citenamefont {Detwiler}, \citenamefont
  {Men\'endez},\ and\ \citenamefont {Vissani}}]{Agostini:2022zub}%
  \BibitemOpen
  \bibfield  {author} {\bibinfo {author} {\bibfnamefont {M.}~\bibnamefont
  {Agostini}}, \bibinfo {author} {\bibfnamefont {G.}~\bibnamefont {Benato}},
  \bibinfo {author} {\bibfnamefont {J.~A.}\ \bibnamefont {Detwiler}}, \bibinfo
  {author} {\bibfnamefont {J.}~\bibnamefont {Men\'endez}}, \ and\ \bibinfo
  {author} {\bibfnamefont {F.}~\bibnamefont {Vissani}},\ }\href {\doibase
  10.1103/RevModPhys.95.025002} {\bibfield  {journal} {\bibinfo  {journal}
  {Rev. Mod. Phys.}\ }\textbf {\bibinfo {volume} {95}},\ \bibinfo {pages}
  {025002} (\bibinfo {year} {2023})},\ \Eprint
  {http://arxiv.org/abs/2202.01787} {arXiv:2202.01787 [hep-ex]} \BibitemShut
  {NoStop}%
\bibitem [{\citenamefont {Liao}\ \emph
  {et~al.}(2020{\natexlab{a}})\citenamefont {Liao}, \citenamefont {Ma},\ and\
  \citenamefont {Wang}}]{Liao:2019gex}%
  \BibitemOpen
  \bibfield  {author} {\bibinfo {author} {\bibfnamefont {Y.}~\bibnamefont
  {Liao}}, \bibinfo {author} {\bibfnamefont {X.-D.}\ \bibnamefont {Ma}}, \ and\
  \bibinfo {author} {\bibfnamefont {H.-L.}\ \bibnamefont {Wang}},\ }\href
  {\doibase 10.1007/JHEP01(2020)127} {\bibfield  {journal} {\bibinfo  {journal}
  {JHEP}\ }\textbf {\bibinfo {volume} {01}},\ \bibinfo {pages} {127} (\bibinfo
  {year} {2020}{\natexlab{a}})},\ \Eprint {http://arxiv.org/abs/1909.06272}
  {arXiv:1909.06272 [hep-ph]} \BibitemShut {NoStop}%
\bibitem [{\citenamefont {Liao}\ \emph
  {et~al.}(2020{\natexlab{b}})\citenamefont {Liao}, \citenamefont {Ma},\ and\
  \citenamefont {Wang}}]{Liao:2020roy}%
  \BibitemOpen
  \bibfield  {author} {\bibinfo {author} {\bibfnamefont {Y.}~\bibnamefont
  {Liao}}, \bibinfo {author} {\bibfnamefont {X.-D.}\ \bibnamefont {Ma}}, \ and\
  \bibinfo {author} {\bibfnamefont {H.-L.}\ \bibnamefont {Wang}},\ }\href
  {\doibase 10.1007/JHEP03(2020)120} {\bibfield  {journal} {\bibinfo  {journal}
  {JHEP}\ }\textbf {\bibinfo {volume} {03}},\ \bibinfo {pages} {120} (\bibinfo
  {year} {2020}{\natexlab{b}})},\ \Eprint {http://arxiv.org/abs/2001.07378}
  {arXiv:2001.07378 [hep-ph]} \BibitemShut {NoStop}%
\bibitem [{\citenamefont {Zhou}(2022)}]{Zhou:2021lnl}%
  \BibitemOpen
  \bibfield  {author} {\bibinfo {author} {\bibfnamefont {G.}~\bibnamefont
  {Zhou}},\ }\href {\doibase 10.1007/JHEP06(2022)127} {\bibfield  {journal}
  {\bibinfo  {journal} {JHEP}\ }\textbf {\bibinfo {volume} {06}},\ \bibinfo
  {pages} {127} (\bibinfo {year} {2022})},\ \Eprint
  {http://arxiv.org/abs/2112.00767} {arXiv:2112.00767 [hep-ph]} \BibitemShut
  {NoStop}%
\bibitem [{\citenamefont {Faessler}\ \emph {et~al.}(1997)\citenamefont
  {Faessler}, \citenamefont {Kovalenko}, \citenamefont {Simkovic},\ and\
  \citenamefont {Schwieger}}]{Faessler:1996ph}%
  \BibitemOpen
  \bibfield  {author} {\bibinfo {author} {\bibfnamefont {A.}~\bibnamefont
  {Faessler}}, \bibinfo {author} {\bibfnamefont {S.}~\bibnamefont {Kovalenko}},
  \bibinfo {author} {\bibfnamefont {F.}~\bibnamefont {Simkovic}}, \ and\
  \bibinfo {author} {\bibfnamefont {J.}~\bibnamefont {Schwieger}},\ }\href
  {\doibase 10.1103/PhysRevLett.78.183} {\bibfield  {journal} {\bibinfo
  {journal} {Phys. Rev. Lett.}\ }\textbf {\bibinfo {volume} {78}},\ \bibinfo
  {pages} {183} (\bibinfo {year} {1997})},\ \Eprint
  {http://arxiv.org/abs/hep-ph/9612357} {arXiv:hep-ph/9612357} \BibitemShut
  {NoStop}%
\bibitem [{\citenamefont {Feng}\ \emph {et~al.}(2019)\citenamefont {Feng},
  \citenamefont {Jin}, \citenamefont {Tuo},\ and\ \citenamefont
  {Xia}}]{Feng:2018pdq}%
  \BibitemOpen
  \bibfield  {author} {\bibinfo {author} {\bibfnamefont {X.}~\bibnamefont
  {Feng}}, \bibinfo {author} {\bibfnamefont {L.-C.}\ \bibnamefont {Jin}},
  \bibinfo {author} {\bibfnamefont {X.-Y.}\ \bibnamefont {Tuo}}, \ and\
  \bibinfo {author} {\bibfnamefont {S.-C.}\ \bibnamefont {Xia}},\ }\href
  {\doibase 10.1103/PhysRevLett.122.022001} {\bibfield  {journal} {\bibinfo
  {journal} {Phys. Rev. Lett.}\ }\textbf {\bibinfo {volume} {122}},\ \bibinfo
  {pages} {022001} (\bibinfo {year} {2019})},\ \Eprint
  {http://arxiv.org/abs/1809.10511} {arXiv:1809.10511 [hep-lat]} \BibitemShut
  {NoStop}%
\bibitem [{\citenamefont {Tuo}\ \emph {et~al.}(2019)\citenamefont {Tuo},
  \citenamefont {Feng},\ and\ \citenamefont {Jin}}]{Tuo:2019bue}%
  \BibitemOpen
  \bibfield  {author} {\bibinfo {author} {\bibfnamefont {X.-Y.}\ \bibnamefont
  {Tuo}}, \bibinfo {author} {\bibfnamefont {X.}~\bibnamefont {Feng}}, \ and\
  \bibinfo {author} {\bibfnamefont {L.-C.}\ \bibnamefont {Jin}},\ }\href
  {\doibase 10.1103/PhysRevD.100.094511} {\bibfield  {journal} {\bibinfo
  {journal} {Phys. Rev. D}\ }\textbf {\bibinfo {volume} {100}},\ \bibinfo
  {pages} {094511} (\bibinfo {year} {2019})},\ \Eprint
  {http://arxiv.org/abs/1909.13525} {arXiv:1909.13525 [hep-lat]} \BibitemShut
  {NoStop}%
\bibitem [{\citenamefont {Nicholson}\ \emph {et~al.}(2018)\citenamefont
  {Nicholson} \emph {et~al.}}]{Nicholson:2018mwc}%
  \BibitemOpen
  \bibfield  {author} {\bibinfo {author} {\bibfnamefont {A.}~\bibnamefont
  {Nicholson}} \emph {et~al.},\ }\href {\doibase
  10.1103/PhysRevLett.121.172501} {\bibfield  {journal} {\bibinfo  {journal}
  {Phys. Rev. Lett.}\ }\textbf {\bibinfo {volume} {121}},\ \bibinfo {pages}
  {172501} (\bibinfo {year} {2018})},\ \Eprint
  {http://arxiv.org/abs/1805.02634} {arXiv:1805.02634 [nucl-th]} \BibitemShut
  {NoStop}%
\bibitem [{\citenamefont {Faessler}\ \emph {et~al.}(2014)\citenamefont
  {Faessler}, \citenamefont {Gonz\'alez}, \citenamefont {Kovalenko},\ and\
  \citenamefont {\v{S}imkovic}}]{Faessler:2014kka}%
  \BibitemOpen
  \bibfield  {author} {\bibinfo {author} {\bibfnamefont {A.}~\bibnamefont
  {Faessler}}, \bibinfo {author} {\bibfnamefont {M.}~\bibnamefont
  {Gonz\'alez}}, \bibinfo {author} {\bibfnamefont {S.}~\bibnamefont
  {Kovalenko}}, \ and\ \bibinfo {author} {\bibfnamefont {F.}~\bibnamefont
  {\v{S}imkovic}},\ }\href {\doibase 10.1103/PhysRevD.90.096010} {\bibfield
  {journal} {\bibinfo  {journal} {Phys. Rev. D}\ }\textbf {\bibinfo {volume}
  {90}},\ \bibinfo {pages} {096010} (\bibinfo {year} {2014})},\ \Eprint
  {http://arxiv.org/abs/1408.6077} {arXiv:1408.6077 [hep-ph]} \BibitemShut
  {NoStop}%
\bibitem [{\citenamefont {Dekens}\ \emph {et~al.}(2020)\citenamefont {Dekens},
  \citenamefont {de~Vries}, \citenamefont {Fuyuto}, \citenamefont
  {Mereghetti},\ and\ \citenamefont {Zhou}}]{Dekens:2020ttz}%
  \BibitemOpen
  \bibfield  {author} {\bibinfo {author} {\bibfnamefont {W.}~\bibnamefont
  {Dekens}}, \bibinfo {author} {\bibfnamefont {J.}~\bibnamefont {de~Vries}},
  \bibinfo {author} {\bibfnamefont {K.}~\bibnamefont {Fuyuto}}, \bibinfo
  {author} {\bibfnamefont {E.}~\bibnamefont {Mereghetti}}, \ and\ \bibinfo
  {author} {\bibfnamefont {G.}~\bibnamefont {Zhou}},\ }\href {\doibase
  10.1007/JHEP06(2020)097} {\bibfield  {journal} {\bibinfo  {journal} {JHEP}\
  }\textbf {\bibinfo {volume} {06}},\ \bibinfo {pages} {097} (\bibinfo {year}
  {2020})},\ \Eprint {http://arxiv.org/abs/2002.07182} {arXiv:2002.07182
  [hep-ph]} \BibitemShut {NoStop}%
\bibitem [{\citenamefont {Fang}\ \emph {et~al.}(2022)\citenamefont {Fang},
  \citenamefont {Li},\ and\ \citenamefont {Zhang}}]{Fang:2021jfv}%
  \BibitemOpen
  \bibfield  {author} {\bibinfo {author} {\bibfnamefont {D.-L.}\ \bibnamefont
  {Fang}}, \bibinfo {author} {\bibfnamefont {Y.-F.}\ \bibnamefont {Li}}, \ and\
  \bibinfo {author} {\bibfnamefont {Y.-Y.}\ \bibnamefont {Zhang}},\ }\href
  {\doibase 10.1016/j.physletb.2022.137346} {\bibfield  {journal} {\bibinfo
  {journal} {Phys. Lett. B}\ }\textbf {\bibinfo {volume} {833}},\ \bibinfo
  {pages} {137346} (\bibinfo {year} {2022})},\ \Eprint
  {http://arxiv.org/abs/2112.12779} {arXiv:2112.12779 [hep-ph]} \BibitemShut
  {NoStop}%
\bibitem [{\citenamefont {Dekens}\ \emph {et~al.}(2023)\citenamefont {Dekens},
  \citenamefont {de~Vries}, \citenamefont {Mereghetti}, \citenamefont
  {Men\'endez}, \citenamefont {Soriano},\ and\ \citenamefont
  {Zhou}}]{Dekens:2023iyc}%
  \BibitemOpen
  \bibfield  {author} {\bibinfo {author} {\bibfnamefont {W.}~\bibnamefont
  {Dekens}}, \bibinfo {author} {\bibfnamefont {J.}~\bibnamefont {de~Vries}},
  \bibinfo {author} {\bibfnamefont {E.}~\bibnamefont {Mereghetti}}, \bibinfo
  {author} {\bibfnamefont {J.}~\bibnamefont {Men\'endez}}, \bibinfo {author}
  {\bibfnamefont {P.}~\bibnamefont {Soriano}}, \ and\ \bibinfo {author}
  {\bibfnamefont {G.}~\bibnamefont {Zhou}},\ }\href {\doibase
  10.1103/PhysRevC.108.045501} {\bibfield  {journal} {\bibinfo  {journal}
  {Phys. Rev. C}\ }\textbf {\bibinfo {volume} {108}},\ \bibinfo {pages}
  {045501} (\bibinfo {year} {2023})},\ \Eprint
  {http://arxiv.org/abs/2303.04168} {arXiv:2303.04168 [hep-ph]} \BibitemShut
  {NoStop}%
\bibitem [{\citenamefont {Fang}\ \emph {et~al.}(2024)\citenamefont {Fang},
  \citenamefont {Li}, \citenamefont {Zhang},\ and\ \citenamefont
  {Zhu}}]{Fang:2024hzy}%
  \BibitemOpen
  \bibfield  {author} {\bibinfo {author} {\bibfnamefont {D.-L.}\ \bibnamefont
  {Fang}}, \bibinfo {author} {\bibfnamefont {Y.-F.}\ \bibnamefont {Li}},
  \bibinfo {author} {\bibfnamefont {Y.-Y.}\ \bibnamefont {Zhang}}, \ and\
  \bibinfo {author} {\bibfnamefont {J.-Y.}\ \bibnamefont {Zhu}},\ }\href
  {\doibase 10.1007/JHEP08(2024)217} {\bibfield  {journal} {\bibinfo  {journal}
  {JHEP}\ }\textbf {\bibinfo {volume} {08}},\ \bibinfo {pages} {217} (\bibinfo
  {year} {2024})},\ \Eprint {http://arxiv.org/abs/2404.12316} {arXiv:2404.12316
  [hep-ph]} \BibitemShut {NoStop}%
\bibitem [{\citenamefont {Dekens}\ \emph {et~al.}(2024)\citenamefont {Dekens},
  \citenamefont {de~Vries}, \citenamefont {Castillo}, \citenamefont
  {Men\'endez}, \citenamefont {Mereghetti}, \citenamefont {Plakkot},
  \citenamefont {Soriano},\ and\ \citenamefont {Zhou}}]{Dekens:2024hlz}%
  \BibitemOpen
  \bibfield  {author} {\bibinfo {author} {\bibfnamefont {W.}~\bibnamefont
  {Dekens}}, \bibinfo {author} {\bibfnamefont {J.}~\bibnamefont {de~Vries}},
  \bibinfo {author} {\bibfnamefont {D.}~\bibnamefont {Castillo}}, \bibinfo
  {author} {\bibfnamefont {J.}~\bibnamefont {Men\'endez}}, \bibinfo {author}
  {\bibfnamefont {E.}~\bibnamefont {Mereghetti}}, \bibinfo {author}
  {\bibfnamefont {V.}~\bibnamefont {Plakkot}}, \bibinfo {author} {\bibfnamefont
  {P.}~\bibnamefont {Soriano}}, \ and\ \bibinfo {author} {\bibfnamefont
  {G.}~\bibnamefont {Zhou}},\ }\href {\doibase 10.1007/JHEP09(2024)201}
  {\bibfield  {journal} {\bibinfo  {journal} {JHEP}\ }\textbf {\bibinfo
  {volume} {09}},\ \bibinfo {pages} {201} (\bibinfo {year} {2024})},\ \Eprint
  {http://arxiv.org/abs/2402.07993} {arXiv:2402.07993 [hep-ph]} \BibitemShut
  {NoStop}%
\bibitem [{\citenamefont {Cirigliano}\ \emph
  {et~al.}(2018{\natexlab{b}})\citenamefont {Cirigliano}, \citenamefont
  {Dekens}, \citenamefont {Mereghetti},\ and\ \citenamefont
  {Walker-Loud}}]{Cirigliano:2017tvr}%
  \BibitemOpen
  \bibfield  {author} {\bibinfo {author} {\bibfnamefont {V.}~\bibnamefont
  {Cirigliano}}, \bibinfo {author} {\bibfnamefont {W.}~\bibnamefont {Dekens}},
  \bibinfo {author} {\bibfnamefont {E.}~\bibnamefont {Mereghetti}}, \ and\
  \bibinfo {author} {\bibfnamefont {A.}~\bibnamefont {Walker-Loud}},\ }\href
  {\doibase 10.1103/PhysRevC.97.065501} {\bibfield  {journal} {\bibinfo
  {journal} {Phys. Rev. C}\ }\textbf {\bibinfo {volume} {97}},\ \bibinfo
  {pages} {065501} (\bibinfo {year} {2018}{\natexlab{b}})},\ \bibinfo {note}
  {[Erratum: Phys.Rev.C 100, 019903 (2019)]},\ \Eprint
  {http://arxiv.org/abs/1710.01729} {arXiv:1710.01729 [hep-ph]} \BibitemShut
  {NoStop}%
\bibitem [{\citenamefont {Pontecorvo}(1957)}]{Pontecorvo:1957qd}%
  \BibitemOpen
  \bibfield  {author} {\bibinfo {author} {\bibfnamefont {B.}~\bibnamefont
  {Pontecorvo}},\ }\href@noop {} {\bibfield  {journal} {\bibinfo  {journal}
  {Zh. Eksp. Teor. Fiz.}\ }\textbf {\bibinfo {volume} {34}},\ \bibinfo {pages}
  {247} (\bibinfo {year} {1957})}\BibitemShut {NoStop}%
\bibitem [{\citenamefont {Maki}\ \emph {et~al.}(1962)\citenamefont {Maki},
  \citenamefont {Nakagawa},\ and\ \citenamefont {Sakata}}]{Maki:1962mu}%
  \BibitemOpen
  \bibfield  {author} {\bibinfo {author} {\bibfnamefont {Z.}~\bibnamefont
  {Maki}}, \bibinfo {author} {\bibfnamefont {M.}~\bibnamefont {Nakagawa}}, \
  and\ \bibinfo {author} {\bibfnamefont {S.}~\bibnamefont {Sakata}},\ }\href
  {\doibase 10.1143/PTP.28.870} {\bibfield  {journal} {\bibinfo  {journal}
  {Prog. Theor. Phys.}\ }\textbf {\bibinfo {volume} {28}},\ \bibinfo {pages}
  {870} (\bibinfo {year} {1962})}\BibitemShut {NoStop}%
\bibitem [{\citenamefont {Doi}\ \emph {et~al.}(1985)\citenamefont {Doi},
  \citenamefont {Kotani},\ and\ \citenamefont {Takasugi}}]{Doi:Masaru}%
  \BibitemOpen
  \bibfield  {author} {\bibinfo {author} {\bibfnamefont {M.}~\bibnamefont
  {Doi}}, \bibinfo {author} {\bibfnamefont {T.}~\bibnamefont {Kotani}}, \ and\
  \bibinfo {author} {\bibfnamefont {E.}~\bibnamefont {Takasugi}},\ }\href
  {\doibase 10.1143/PTPS.83.1} {\bibfield  {journal} {\bibinfo  {journal}
  {Progress of Theoretical Physics Supplement}\ }\textbf {\bibinfo {volume}
  {83}},\ \bibinfo {pages} {1} (\bibinfo {year} {1985})}\BibitemShut {NoStop}%
\bibitem [{\citenamefont {Lindner}\ \emph {et~al.}(2011)\citenamefont
  {Lindner}, \citenamefont {Merle},\ and\ \citenamefont
  {Niro}}]{Lindner:2010wr}%
  \BibitemOpen
  \bibfield  {author} {\bibinfo {author} {\bibfnamefont {M.}~\bibnamefont
  {Lindner}}, \bibinfo {author} {\bibfnamefont {A.}~\bibnamefont {Merle}}, \
  and\ \bibinfo {author} {\bibfnamefont {V.}~\bibnamefont {Niro}},\ }\href
  {\doibase 10.1088/1475-7516/2011/01/034} {\bibfield  {journal} {\bibinfo
  {journal} {JCAP}\ }\textbf {\bibinfo {volume} {01}},\ \bibinfo {pages} {034}
  (\bibinfo {year} {2011})},\ \bibinfo {note} {[Erratum: JCAP 07, E01
  (2014)]},\ \Eprint {http://arxiv.org/abs/1011.4950} {arXiv:1011.4950
  [hep-ph]} \BibitemShut {NoStop}%
\bibitem [{\citenamefont {Tuo}\ \emph {et~al.}(2022)\citenamefont {Tuo},
  \citenamefont {Feng},\ and\ \citenamefont {Jin}}]{Tuo:2022hft}%
  \BibitemOpen
  \bibfield  {author} {\bibinfo {author} {\bibfnamefont {X.-Y.}\ \bibnamefont
  {Tuo}}, \bibinfo {author} {\bibfnamefont {X.}~\bibnamefont {Feng}}, \ and\
  \bibinfo {author} {\bibfnamefont {L.-C.}\ \bibnamefont {Jin}},\ }\href
  {\doibase 10.1103/PhysRevD.106.074510} {\bibfield  {journal} {\bibinfo
  {journal} {Phys. Rev. D}\ }\textbf {\bibinfo {volume} {106}},\ \bibinfo
  {pages} {074510} (\bibinfo {year} {2022})},\ \Eprint
  {http://arxiv.org/abs/2206.00879} {arXiv:2206.00879 [hep-lat]} \BibitemShut
  {NoStop}%
\bibitem [{\citenamefont {Barea}\ \emph {et~al.}(2015)\citenamefont {Barea},
  \citenamefont {Kotila},\ and\ \citenamefont {Iachello}}]{Barea:2015zfa}%
  \BibitemOpen
  \bibfield  {author} {\bibinfo {author} {\bibfnamefont {J.}~\bibnamefont
  {Barea}}, \bibinfo {author} {\bibfnamefont {J.}~\bibnamefont {Kotila}}, \
  and\ \bibinfo {author} {\bibfnamefont {F.}~\bibnamefont {Iachello}},\ }\href
  {\doibase 10.1103/PhysRevD.92.093001} {\bibfield  {journal} {\bibinfo
  {journal} {Phys. Rev. D}\ }\textbf {\bibinfo {volume} {92}},\ \bibinfo
  {pages} {093001} (\bibinfo {year} {2015})},\ \Eprint
  {http://arxiv.org/abs/1509.01925} {arXiv:1509.01925 [hep-ph]} \BibitemShut
  {NoStop}%
\bibitem [{\citenamefont {Asaka}\ \emph {et~al.}(2016)\citenamefont {Asaka},
  \citenamefont {Eijima},\ and\ \citenamefont {Ishida}}]{Asaka:2016zib}%
  \BibitemOpen
  \bibfield  {author} {\bibinfo {author} {\bibfnamefont {T.}~\bibnamefont
  {Asaka}}, \bibinfo {author} {\bibfnamefont {S.}~\bibnamefont {Eijima}}, \
  and\ \bibinfo {author} {\bibfnamefont {H.}~\bibnamefont {Ishida}},\ }\href
  {\doibase 10.1016/j.physletb.2016.09.044} {\bibfield  {journal} {\bibinfo
  {journal} {Phys. Lett. B}\ }\textbf {\bibinfo {volume} {762}},\ \bibinfo
  {pages} {371} (\bibinfo {year} {2016})},\ \Eprint
  {http://arxiv.org/abs/1606.06686} {arXiv:1606.06686 [hep-ph]} \BibitemShut
  {NoStop}%
\bibitem [{\citenamefont {Babi\v{c}}\ \emph {et~al.}(2018)\citenamefont
  {Babi\v{c}}, \citenamefont {Kovalenko}, \citenamefont {Krivoruchenko},\ and\
  \citenamefont {\v{S}imkovic}}]{Babic:2018ikc}%
  \BibitemOpen
  \bibfield  {author} {\bibinfo {author} {\bibfnamefont {A.}~\bibnamefont
  {Babi\v{c}}}, \bibinfo {author} {\bibfnamefont {S.}~\bibnamefont
  {Kovalenko}}, \bibinfo {author} {\bibfnamefont {M.~I.}\ \bibnamefont
  {Krivoruchenko}}, \ and\ \bibinfo {author} {\bibfnamefont {F.}~\bibnamefont
  {\v{S}imkovic}},\ }\href {\doibase 10.1103/PhysRevD.98.015003} {\bibfield
  {journal} {\bibinfo  {journal} {Phys. Rev. D}\ }\textbf {\bibinfo {volume}
  {98}},\ \bibinfo {pages} {015003} (\bibinfo {year} {2018})},\ \Eprint
  {http://arxiv.org/abs/1804.04218} {arXiv:1804.04218 [hep-ph]} \BibitemShut
  {NoStop}%
\bibitem [{\citenamefont {Asaka}\ \emph {et~al.}(2021)\citenamefont {Asaka},
  \citenamefont {Ishida},\ and\ \citenamefont {Tanaka}}]{Asaka:2020wfo}%
  \BibitemOpen
  \bibfield  {author} {\bibinfo {author} {\bibfnamefont {T.}~\bibnamefont
  {Asaka}}, \bibinfo {author} {\bibfnamefont {H.}~\bibnamefont {Ishida}}, \
  and\ \bibinfo {author} {\bibfnamefont {K.}~\bibnamefont {Tanaka}},\ }\href
  {\doibase 10.1103/PhysRevD.103.015014} {\bibfield  {journal} {\bibinfo
  {journal} {Phys. Rev. D}\ }\textbf {\bibinfo {volume} {103}},\ \bibinfo
  {pages} {015014} (\bibinfo {year} {2021})},\ \Eprint
  {http://arxiv.org/abs/2012.12564} {arXiv:2012.12564 [hep-ph]} \BibitemShut
  {NoStop}%
\bibitem [{\citenamefont {Mehen}\ and\ \citenamefont
  {Stewart}(1999)}]{Mehen:1998tp}%
  \BibitemOpen
  \bibfield  {author} {\bibinfo {author} {\bibfnamefont {T.}~\bibnamefont
  {Mehen}}\ and\ \bibinfo {author} {\bibfnamefont {I.~W.}\ \bibnamefont
  {Stewart}},\ }\href {\doibase 10.1103/PhysRevC.59.2365} {\bibfield  {journal}
  {\bibinfo  {journal} {Phys. Rev. C}\ }\textbf {\bibinfo {volume} {59}},\
  \bibinfo {pages} {2365} (\bibinfo {year} {1999})},\ \Eprint
  {http://arxiv.org/abs/nucl-th/9809095} {arXiv:nucl-th/9809095} \BibitemShut
  {NoStop}%
\end{thebibliography}%

\end{document}